\documentstyle[12pt,aaspp4,psfig]{article}

\lefthead{S. L. Ellison et. al.}
\righthead{Metal Abundances in the Ly$\alpha$ Forest}

\begin{document}
\newcommand{\lya}{Ly$\alpha$}
\newcommand{\lyb}{Ly$\beta$}
\newcommand{\lyg}{Ly$\gamma$}
\newcommand{\HI}{H~\textsc{I}}
\newcommand{\CIV}{C~\textsc{IV}}
\newcommand{\SiIV}{Si~\textsc{IV}}
\newcommand{\kms}{km s$^{-1}$} 
\def\ltsima{$\; \buildrel < \over \sim \;$}
\def\simlt{\lower.5ex\hbox{\ltsima}}
\def\gtsima{$\; \buildrel > \over \sim \;$}
\def\simgt{\lower.5ex\hbox{\gtsima}}

\title{HIRES Spectroscopy of APM 08279+5525:
Metal Abundances in the Ly$\alpha$ Forest \footnote{The data
presented herein were obtained at the W. M. Keck Observatory, which is
operated as a scientific partnership among the California Institute of
Technology, the University of California and the National Aeronautics
and Space Administration.  The Observatory was made possible by the
generous financial support of the W. M. Keck Foundation.}}

\author{Sara L. Ellison\altaffilmark{2}, Geraint F. Lewis\altaffilmark{3}, Max 
Pettini\altaffilmark{2}, Wallace L. W. Sargent\altaffilmark{4},
Frederic H. Chaffee\altaffilmark{5}, Mike J. Irwin\altaffilmark{2}}

\altaffiltext{2}{
Institute of Astronomy, Madingley Road, Cambridge CB3 0HA, UK \nl
Electronic mail contact: {\tt sara@ast.cam.ac.uk}}

\altaffiltext{3}{ Fellow of the Pacific Institute of Mathematical
Sciences 1998-1999, \nl Dept. of Physics and Astronomy, University of
Victoria, PO Box 3055, Victoria, B.C., V8W 3P6, Canada \nl \&
Astronomy Dept., University of Washington, Box 351580, Seattle,
WA 98195-1580}

\altaffiltext{4}{Palomar Observatory, Caltech 105--24, Pasadena, CA 91125}

\altaffiltext{5}{ W. M. Keck Observatory, 65--1120 Mamalahoa Hwy,
Kamuela, HI 96743}



\begin{abstract}

We present high S/N echelle spectra of the
recently discovered ultraluminous QSO APM~08279+5255 and use
these data to re-examine the abundance of Carbon in
\lya\ forest clouds. In
agreement with previous work, we find that approximately 50\% of
\lya\ clouds with hydrogen column densities log~$N$(H~I)$ \geq
14.5$ have associated weak C~IV absorption with log~$N$(C~IV)$
\simgt 12$, and derive a median $N$(C~IV)/$N$(H~I)$ = 1.4 \times
10^{-3}$. The agreement with earlier estimates of this ratio may
be somewhat fortuitous, however, because we show that previous
analyses have probably overestimated the number of \lya\ clouds
which should be included in this statistic.  

We then investigate whether there is any \CIV\ absorption
associated with weaker \HI\ column densities by stacking 51 C~IV
regions corresponding to 51 \lya\ lines with $13.5 \leq {\rm
log}N{\rm (H~I)} \leq 14.0$. The co-added spectrum has S/N
$\simeq 580$ but shows no composite C~IV absorption. In order to
understand the significance of this non-detection we have stacked
together 51 theoretical C~IV~$\lambda 1548$ lines with individual
values of column density and velocity dispersion scaled
appropriately from the values {\it measured} in the corresponding
\lya\ lines. We find that even if the typical value 
$N$(C~IV)/$N$(H~I)$ = 1.4 \times 10^{-3}$ applies to these lower
column density clouds, the corresponding signal in the stacked
C~IV region is smeared by the likely random difference in
redshift between C~IV and \lya\ absorption and becomes very
difficult to recognize. This seems to be a fundamental limitation
of the stacking method which may well explain why in the past it
has led to underestimates of the metallicity of the \lya\ forest.

We also analyze our spectra with the pixel-by-pixel optical depth
technique recently developed by 
Cowie \& Songaila (1998) and find evidence for
net C~IV absorption in \lya\ clouds with optical depths as low as 
$\tau {\rm (Ly\alpha)} = 0.5 - 2$, as these authors did.
However, we show with simulations that even this 
method requires higher sensitivities than reached up to now to 
be confident that the ratio 
$N$(C~IV)/$N$(H~I) remains constant down to column densities
below log~$N$(H~I)$ \simeq 14.0$\,.
We conclude that the 
question of whether there is a uniform degree of metal enrichment in the 
\lya\ forest at all column densities has yet to be fully answered.
Future progress in this area will probably require
concerted efforts to push further the detection limit for 
C~IV lines in selected bright QSOs.

\end{abstract}


\keywords{galaxies: formation -- galaxies: intergalactic medium --
quasars: absorption lines -- quasars: individuals (APM 08279+5255)}


%

\section{Introduction}\label{introduction}

Nearly thirty years after its discovery by Lynds (1971), the \lya\
forest continues to provide fertile ground for observational and
theoretical studies.  New echelle spectrographs on large
telescopes, most notably the High Resolution
Spectrograph (HIRES, \cite{v94}) on Keck I, have provided
spectra of exceptional quality and detail.
This has permitted extensive study of many important properties of the
\lya\ forest as comprehensively reviewed by Rauch (1998).
Complementary to this quality observational data, cosmological simulations
which utilize gas hydrodynamics have improved our understanding of the
nature of the forest clouds.  These simulations have shown that the
formation of the \lya\ forest is a natural consequence of the growth
of structure in the universe through hierarchical clustering in the
presence of a UV ionizing background (e.g. \cite{cmo94}; \cite{pmk95};
\cite{hkw96}; \cite{bd97}).  In this class of models, low column density \HI\
clouds (log~$N$(\HI)$<$14, where $N$(\HI) is measured in cm$^{-2}$)
are found preferentially in voids whilst
stronger lines are associated with higher density clouds which arise in
filamentary structures around collapsed objects.

Important clues to the origin of \lya\ clouds can be gleaned from
their clustering properties and chemical composition.
Early investigations of the \lya\ forest
found no evidence of metals associated with the
H~\textsc{I} clouds, suggesting that they consisted of pristine
material (\cite{syb80}).
More concerted efforts however, did provide some evidence for metals
in the forest  either individually (Meyer \& York 1987) or in a
stacked spectrum (Lu 1991).  The realization that metal
enrichment is in fact widespread in high column density \lya\ clouds has come
relatively recently with the availability of HIRES on the Keck I telescope.  
A number of studies (e.g. \cite{t95}; \cite{csk95}; \cite{sc96}) have
found that C~\textsc{IV} absorption is seen in approximately
50\% of forest clouds with log~$N$(H~\textsc{I})$>$14.5 and in about 90\%
of clouds with log~$N$(H~\textsc{I})$>$15.
Evidently the intergalactic medium (IGM) has been enriched by the
products of stellar nucleosynthesis even at redshifts as high as $z
\sim 3.5$.  Photoionization models (e.g. \cite{hdh97}) 
can reproduce the observed
\CIV/\HI\ ratio at $z \sim 3$ for a typical 
Carbon abundance\footnote{$\left[ {X \over H}
\right] = \log \left( {N(X) \over N(H)} \right) - \log \left( {N(X)
\over N(H)} \right) _{\odot}$} [C/H] = $-2.5$\,.
Rauch, Haehnelt, \& Steinmetz (1997) predict an order
of magnitude scatter in [C/H] in models where random sight-lines are
cast through protogalactic clumps, a prediction corroborated by the data. 

Two scenarios have been suggested to explain the presence of
metals in the \lya\ forest; early pre-enrichment by a widespread
episode of star formation, possibly associated with Pop III stars,
or {\it in-situ} enrichment whereby the \HI\ cloud 
is enriched locally either by contamination from a nearby galaxy or by
star formation within the cloud itself (\cite{og96}; \cite{go97}).  
Cosmological simulations of a
two phase ISM recently performed by Gnedin (1998) confirm the results of
Gnedin \& Ostriker (1997) that the dominant process for transportation of heavy
elements into the IGM is
mergers of protogalaxies in dark matter halos.
One may expect that for the scenario of {\it in-situ} enrichment,
where metals are synthesized relatively close to the \lya\ clouds being
observed, the metallicities would be highly inhomogeneous depending on the
proximity of a given \lya\ cloud
to a star-forming region and its past merger history.  If, however,
the bulk of the metals is formed by an episode of Pop III star
formation at sufficiently high redshift ($z \sim 14$ in the models of
Ostriker \& Gnedin 1996), a more uniform [C/H] may have been established
by $z \sim 3$, if the mixing mechanism is efficient.

Discrimination between these two enrichment scenarios is probably best
addressed in the low column density regime.  Since the optical
depth of \lya\ clouds is roughly indicative of the baryon
overdensity, studying low column density lines will give us an
insight into the low mass systems.  
In the simulations of Gnedin \& Ostriker (1997), for
example, a sudden drop in metallicity is predicted for \lya\ clouds
with log~$N$(\HI)$<$14 where the star formation rate is expected to
be much lower than in higher density clouds.   
However, determining the [C/H] abundance in weak Ly$\alpha$ clouds
has been observationally challenging due to the extreme weakness of
C~\textsc{IV} absorption when log~N(H~\textsc{I}) $\simlt 14$.
In order to overcome the
difficulty of a direct C~\textsc{IV} detection, this problem has recently
been tackled with two different approaches  which have
produced apparently conflicting results.  

Lu et al. (1998) tackled the problem by stacking almost 300
\CIV\ regions to produce a composite spectrum, a method first
applied to QSO spectra by Norris,  Peterson \& Hartwick (1983).  Having
found no \CIV\ in their summed spectrum (which had a final S/N ratio of
1860) Lu et al. used Monte Carlo simulations to place an upper
limit on the Carbon abundance in clouds with log~$N$(\HI)$<14$ of
[C/H]$<-3.5$.  
A different conclusion was reached by Cowie \& Songaila
(1998) who used distributions of Ly$\alpha$ optical depths,
$\tau {\rm (Ly\alpha)}$, to build 
distributions of corresponding $\tau$(\CIV) on a
pixel-by-pixel basis.  A comparison  between the $\tau$(\CIV)
distribution and a reference  `blank' distribution
showed a residual signal sufficiently strong to be consistent with  
[C/H]$\simeq -2.5$  for \lya\ clouds with log~$N$(H~\textsc{I}) as
low as 13.5.  
Thus, whilst Lu et al. interpreted their results as evidence against
a uniform enrichment of the IGM at $z \sim 3$, 
Cowie \& Songaila  proposed that
transportation/ejection mechanisms from early sites of star formation
are much more efficient than anticipated, in order to pollute
\lya\ clouds uniformly over a range of N(\HI) of nearly 4 orders of
magnitude.

In this paper we aim to address the contradiction between these two
earlier analyses by taking another look at the metallicity in the
\lya\ forest using new data which are among the best ever obtained for
this purpose; almost 9 hours of HIRES observations
(described in \S 2) of the ultra-luminous
QSO APM 08279+5255.  In \S 3 we define the redshift interval over which we
search for \CIV\ absorption;
the very high S/N of the spectrum in the \CIV\
region allows us to 
refine earlier estimates of the typical \CIV/\HI\ ratio in
\lya\ lines with log~$N$(\HI)$> 14.5$ (\S 4).  In \S 5 we
investigate whether this high quality, independent data set can help
resolve the current discrepancy between the results of Lu et
al. (1998) and Cowie \& Songaila (1998), thus shedding some light on the
origin of the metals observed in low column density \lya\ clouds. 
We summarize our main results in \S 6. 

\section{Observations and Data Reduction}

The target for these observations is APM 08279+5255, an ultra-luminous
Broad Absorption Line (BAL) quasar discovered by Irwin et al. (1998)
during a survey of Galactic halo carbon stars.  The emission redshift 
$z_{\rm em} = 3.87$ measured by Irwin et al. from the 
\CIV\ $\lambda 1549$ and N~\textsc{V} $\lambda 1240$ 
emission lines has recently been refined to 
$z_{\rm em} = 3.9110$ by Downes et al. (1999) who detected 
CO emission. Given the difficulty in measuring a precise redshift
from the UV emission lines which are affected by the BAL phemomenon,
the difference is probably not significant; in the following analysis we
adopt $z = 3.9110$  as the systemic redshift of the QSO.
Positionally coincident with an IRAS Faint Source Catalog
object with a 60 micron flux of 0.51Jy, APM 08279+5255 has an
optical R magnitude=15.2 and an
inferred bolometric luminosity of $5 \times
10^{15}$L$_{\odot}$ making it the most luminous object currently known.
Adaptive optics imaging with the CFHT (\cite{ltp98}) has resolved APM
08279+5255 into two images separated by 0.35 arcsec in the NE--SW
direction with an intensity ratio $I_{NE}$/$I_{SW} = 1.2 \pm
0.25$ in the $H$-band.

For our purposes, APM 08279+5255 is a nearly ideal background source
for investigating the \lya\ forest and associated metal absorption
lines at high S/N ratio and resolution.
The data presented here were
obtained with HIRES on the Keck I telescope 
on three runs in April and May 1998.
Details of the observations are presented in Table 1.   APM 08279+5255
is unresolved in these observations.  The cross
disperser  and echelle angles were used in a variety of settings 
to give almost complete  wavelength coverage  between 4400 and
9250 \AA. The emission spectrum of a Th-Ar hollow cathode lamp
provided a wavelength reference and a continuum source internal to the
spectrograph was used for a first order correction of the echelle blaze
function. 

The data were reduced using  Barlow's (1999, in preparation) 
customized HIRES reduction
package (HAR).  The individual, sky-subtracted spectra were mapped
onto a linear wavelength scale with a
dispersion  of 0.04 \AA\ per wavelength bin and then co-added
with a weight
proportional to their S/N.  
Finally, this co-added spectrum was normalized
by fitting a cubic spline function with STARLINK software
to continuum regions deemed to be free of absorption.  Even in the
dense forest of \lya\ lines at wavelengths below that of \lya\ emission,
continuum
windows can be identified at the high resolution of our data.

The final co-added spectrum has a resolution of 6 \kms,  
sampled with $\sim$ 3.5 wavelength bins, 
and S/N between 30 and 150. Over the
range of interest for \CIV\ absorption (see \S3 below) 
the typical signal-to-noise ratio is S/N $\simeq 80$. 
At the mean $z_{\rm abs} = 3.41$ the $5 \sigma$ detection
limit for the rest-frame equivalent width of C~IV absorption line
with the typical FWHM = 22~km~s$^{-1}$ (see later) is
$W_0$(1548)$ = 3$~m\AA, which corresponds to a column density
log~$N$(\CIV) = 11.88\,. Thus these data are comparable to those
obtained by Songaila \& Cowie (1996) for Q0014+813 and
Q1422+231, their best observed QSOs.

The full spectrum is described elsewhere (Ellison et al. 1999)
and is available via anonymous ftp from ftp.ast.cam.ac.uk
(pub/sara/APM0827).

\section{Sample Definition}

APM 08279+5255 belongs to a class of objects whose spectra are
known to exhibit broad, high velocity, intrinsic absorption lines.
Before proceeding we must therefore define a working wavelength
interval over which we can study \CIV\ and \lya\ lines in our spectrum
and be confident that we are not confusing them with absorption from
ejected material.  
The upper limit of this interval (for \CIV) is 7280~\AA\  ($z_{\rm max} =
3.701$), at the blue edge of the broad C~IV absorption trough which
corresponds to an ejection velocity $v_{\rm ej} \simeq
13 100$~km~s$^{-1}$ relative to the systemic redshift. The lower
limit for the C IV interval is 6365~\AA\ ($z_{\rm min} = 3.109$),
the wavelength of
\lyb\ emission.  We excluded the region of the spectrum
blueward of this wavelength in order to avoid confusion of \lya\
with \lyb\ lines. We further excluded a small region between 6860 and
6950~\AA\ which is contaminated by the atmospheric B band.






In principle, some of the C~IV doublets in our sample could still
be related to the BAL phenomenon and be intrinsic to the QSO
rather than intervening absorbers, since ejecta have been observed at
velocities as large as 60,000 \kms. However, we found no evidence
of this in APM 08279+5255. For example, none of the C~IV systems at 
$z_{\rm abs} < z_{\rm max}$ show any indication of partial coverage of the QSO
(one of the signatures of intrinsic absorption), 
whereas we do see such an effect for
systems at $z_{\rm abs} > z_{\rm max}$. Most importantly, as we will
discuss below, the density of \CIV\ absorbers per unit redshift
between $z_{\rm min}$ and $z_{\rm max}$ is entirely compatible
with that measured towards non-BAL QSOs---there is no excess of
\CIV\ absorbers within our working wavelength interval.

\section{\CIV\ In \lya\ Clouds with log $N$(\HI) $\geq 14.5$}

Before addressing the question of CIV absorption in \lya\ clouds with 
13.5$ \leq$ log N(\HI) $\leq 14.0$, the first stage of our analysis
is to determine the typical $N$(\CIV)/$N$(\HI) ratio in \lya\ clouds
with log $N$(H~I) $\geq 14.5$.  The aim is to provide an independent
measure of this quantity, and therefore of the metallicity of \lya\ clouds, for
comparison with the results of Cowie et al. (1995) and Songaila
\& Cowie (1996).

These authors found that at a detection limit log~$N$(C~IV) $\geq
12.0$ approximately half of the \lya\ lines with
log~$N$(H~I)$\geq 14.5$ have associated C~IV absorption. \lya\
lines above this column density cut-off are saturated and therefore
not on the linear part of the curve of growth.  Therefore, an
accurate $N$(H~I) cannot normally be measured without higher-order Lyman
lines such as \lyb\ or \lyg\ that are not saturated. 
Cowie et al. (1995) circumvented this difficulty by
assuming that at log~$N$(H~I)$\geq 14.5$ the residual flux in the
core of a \lya\ absorption line is $r_f < 0.025$ 
for the mean value of the Doppler
parameter of $b = 34 $~\kms\ 
determined by Carswell et al. (1991)\footnote{As usual $b= \sqrt 2
\sigma$, where $\sigma$ is the one-dimensional velocity dispersion of
the absorbers projected along the line of sight}.
This approach is of course 
critically sensitive to a precise determination of the 
zero level.
 
Figure 1 shows
the \lya\ forest in APM~08279+5525 between $z_{\rm abs} = 3.109$
and 3.701;  in this redshift interval there are 58 \lya\ lines 
with $r_f < 0.025$.  However a search for corresponding \CIV\
absorption yielded only 22 \CIV\ systems (38\%) 
at a significance level of
$\geq 5\sigma$\,.  

This apparent discrepancy led us to examine how accurately the $r_f
< 0.025$ cut actually selects \lya\ clouds with log~$N$(H~I)$\geq
14.5$. To this end we used the profile fitting package VPFIT
(\cite{w87}) to determine the column density, redshift and
$b$ parameter of \lya\ lines with $r_f <
0.025$.  We found that although these lines are
saturated, in most cases there is sufficient information in the
absorption profiles for VPFIT to converge to a reliable solution.
Fitting all lines which do not have a flat core at zero residual
intensity, showed that 22 out of 58 \lya\ lines with $r_f <
0.025$ do in fact have log~$N$(H~I)$ < 14.5$. We have
reproduced some examples in Figure 2 and Table 3. The revised sample then
consists of 36 \lya\ lines with {\em measured} log~$N$(H~I)$ >
14.5$; 20 of these (56\%) have associated C~IV systems. This
fraction is in good agreement, although somewhat fortuitously,
with the value of $58\pm8$\% reported by Songaila \& Cowie
(1996).

All of the C~\textsc{IV} systems found within our redshift
interval were fitted with Voigt profiles using VPFIT (all \CIV\
lines were unsaturated, thus allowing accurate determinations of their
column densities) and the model parameter fits for $z$, $N$(\CIV), and
$b$ are listed in Table 2.    
Fig 3 is an atlas of the 23 \CIV\ systems whose
corresponding \lya\ lines lie
within our redshift interval; 22 of these have \lya\ with $r_f < 0.025$. 
Only the
20 \CIV\ systems with a measured log~$N$(\HI)$ > 14.5$ are considered in the
statistics below.

In order to derive the typical $N$(C~IV)/$N$(H~I) ratio in \lya\
clouds with log~$N$(H~I) $> 14.5$, we follow the procedure used
by Cowie et al. (1995). For an H~I column density distribution of the
form
\begin{equation}
n(N)dN \propto N^{-1-\beta}dN
\end{equation}
with $\beta \simeq 0.7$ (\cite{pwr93}), the median
$N$(\HI) is $2^{\beta}$ times the minimum value or, in our case,
$5 \times 10^{14}$~cm$^{-2}$. Our $5 \sigma$ detection limit for
C~IV$\lambda 1548$ corresponds to $N$(C~IV)$ = 7 \times
10^{11}$~cm$^{-2}$ for the median $b(C~IV) = 13.4$ \kms\
(see Table 2). Since at this detection level approximately half
of the \lya\ clouds have associated C~IV absorption, we deduce a
median $N$(C~IV)/$N$(H~I)$ \simeq 1.4 \times 10^{-3}$. For
comparison, Songaila \& Cowie (1996) reported median values of
$N$(C~IV)/$N$(H~I) between $1.6 \times 10^{-3}$ and $2.8 \times
10^{-3}$.

  
In Figure 4 we show the column density distribution of C~IV lines
again assumed to be a power law of the form
\begin{equation}
f(N) dN = BN^{\alpha} dN
\end{equation}
where $f(N)$ is the number of systems per column density interval
per unit redshift path. 
The redshift path (used instead of $z$ in
order to account for comoving distances) is given by $X(z)={1
\over 2} [(1+z)^2-1]$ for $q_0 = 0$.\footnote{$q_0$=0 and $H_0 =
65$ km~s$^{-1}$ Mpc$^{-1}$ are adopted throughout this paper
unless otherwise stated.} 
A maximum likelihood (e.g. Schechter \&
Press 1976) fit to the total (unbinned) sample of 20 \CIV\ systems
yielded $\alpha = -1.0 \pm 0.1$ for column densities in the range
$11.74 \leq $log~$N$(\CIV) $\leq 14.25$.  
This value of $\alpha$, which is shown as a solid
line in Figure 4, would indicate a flatter power law distribution
than the value $\alpha = -1.5$ (dashed line in Figure 4) deduced
by Songaila (1997) from her analysis of a larger sample of 81
C~IV absorbers towards seven QSOs. 
Formally, the difference is significant.
A Kolmogorov-Smirnov (K-S)
test returns a 76\% probability that our data points are drawn from
a distribution with $\alpha = -1.0$, but only a $< 1$\% probability that they
arise by chance from a distribution with $\alpha = -1.5$.
However, there are significant variations in the statistics of C~IV absorbers 
between different sight-lines in Songaila's sample and
the integral of our power-law distribution, $\sum
N(C~IV)/\Delta X = 2.3 \times 10^{14}$~cm$^{-2}$,
is the same as that reported by Songaila for one of the sight-lines in 
her study, towards Q0956+122. Thus the difference between our best 
fitting value of $\alpha$ and that determined by Songaila (1997)
may just be due to the limited statistics of the current samples.

\section{\lya\ Clouds with 13.5 $<$ log~$N$(H~\textsc{I})$<$ 14.0}

\subsection{The stacking method}

We now turn to the question of whether \lya\ clouds with 
log~$N$(H~I)$ < 14.0$ have a significantly lower metallicity than
those with log~$N$(H~I)$ > 14.5$. 
Lu et al. (1998) tackled
this problem by stacking together the C~IV regions corresponding
to $\sim 300$ \lya\ lines towards nine QSOs; no significant
signal was found in the final composite spectrum. In constructing
their sample, Lu et al. targeted column densities in the range
$13.5 < $ log$~N(H~I) < 14.0$ and, for expediency, assumed
that this range corresponds to values of residual flux $0.05 <
r_f < 0.38$.
Although our statistics are more limited, we can improve on
previous analyses by constructing our sample more carefully and
by simulating the results of the stacking process for our {\em
observed} distributions of column densities and $b$-values, as we
now discuss.

We started by using VPFIT to fit all unsaturated \lya\ lines in the
spectrum of APM~08279+5525 between $z_{abs} = 3.109$ and
3.701\,.  This produced a sample of 86 lines with 
$13.5 < $log $N(H~I) < 14.0$. 
Had we selected on the basis of $r_f$
rather than $N$(H~I), approximately 30\% of lines in the desired
column density range would have been missed, and 35\% of lines
outside the range would have been erroneously included.
The lines missed are mainly lines with large values of $b$,
and lines which are partially blended but for which VPFIT can
still satisfactorily recover the individual values of $N$(\HI).
In the next step we visually inspected the C~IV region
corresponding to each \lya\ line to assess whether it is suitable
for stacking; regions contaminated by atmospheric features or
blended with other lines were excluded. This produced a final
list of 51 C~IV$\lambda 1548$ regions and 40 \CIV\ $\lambda 1550$
regions each 7~\AA\ wide, which
could be stacked. After reducing each portion of the spectrum to
the rest frame using the values of $z_{abs}$ appropriate to
each \lya\ line (as measured with VPFIT), the regions were summed
and renormalized. The final co-added spectrum has S/N = 580 and
is reproduced in Figure 5a.

It is clear from Figure 5a that no absorption is detected in the
stacked spectrum. In order to assess the significance of this
non-detection, we compare the stacked spectrum with that obtained
by adding together the same number of \CIV\ lines with strengths 
predicted for different values of $N$(\CIV)/$N$(\HI). In
performing these simulations we again made use of the available
information on the distribution of column densities and
$b$-values appropriate to each line, rather than assuming
representative values of these quantities, as was done in earlier
analyses. Specifically, we simulated the absorption profiles of
51 C~IV$\lambda 1548$ lines assigning to each a column density
$N$(\CIV)$_i$ = $N$(\HI)$_i \times A$, where $N$(\HI)$_i$ is the
neutral hydrogen column density returned by VPFIT for the $i$th
\lya\ line, and $A$ is the adopted ratio $N$(C~IV)/$N$(H~I). We
repeated the simulations for three values of $A = 2.5 \times
10^{-3}$ (Cowie \& Songaila 1998), $1.4 \times 10^{-3}$, the
median value deduced here for clouds with log~$N$(H~I)$ \geq
14.5$, and $0.7 \times 10^{-3}$, a factor of two lower than our
median.

Concerning the value of $b$ to be assigned to each C~IV line in
the simulations, we could in principle consider two
possibilities. If $b$ reflected primarily large-scale motions of
the absorbing gas, $b(C~IV)_i = b(H~I)_i$. On the other
hand, if $b$ has mainly a thermal origin, the $b$-values would
scale as the square root of the atomic mass and $b(C~IV)_i =
1/\sqrt{12} \times b(H~I)_i$. The real situation will be
somewhere between these two extremes. Rauch et al. (1996) found
mean and median values of $b(C~IV)$ near 10~\kms\ from
their analysis of 208 C~IV absorption components, and concluded
that both thermal motions and bulk motions contribute to the line
broadening. Our distribution of $b(C~IV)$ for \lya\ clouds
with log~$N$(\HI)$ \geq 14.5$ has a median value $b(C~IV) =
13.4$~\kms\, whereas for $13.5 \leq$ log~$N$(H~I)$ \leq
14.0$ we find a median $b(H~I) = 28.7$. In the simulations
we therefore adopted $b(C~IV)_i \simeq 1/2~b(H~I)_i$,
with the implicit assumption that there is no significant change
in the distribution of $b$-values with column density.

Theoretical C~IV$\lambda 1548$ absorption line profiles were
computed in this manner for the 51 \lya\ lines in our sample,
convolved with the instrumental resolution, and then co-added.
The resulting composite profiles, degraded with random noise
corresponding to the S/N = 580 of the real stacked data, are
shown in Figures 5b, c, and d.
Also shown in these panels is the corresponding theoretical
absorption profile for the mean $N$(\CIV).

The case $A = 2.5 \times 10^{-3}$ (Figure 5b) produces a
composite absorption feature which is significant at the $7
\sigma$ level; such a high value of the $N$(C~IV)/$N$(H~I) ratio
would therefore appear to be excluded by our observations. At
these low optical depths the equivalent width of the co-added
C~IV$\lambda 1548$ line scales linearly with $A$. Thus, if the
median $A = 1.4 \times 10^{-3}$ found above for \lya\ clouds with
log~$N$(\HI) $\geq 14.5$ also applied at log~$N$(\HI) $= 13.5 -
14.0$, we may also expect a detectable signal ($4 \sigma$, Figure
5c), whereas no composite absorption would be recognized if $A =
0.7 \times 10^{-3}$ (Figure 5d).

In the above simulations we have assumed that $z(C~IV)_i =
z(H~I)_i$. However, in reality there is likely to be a
dispersion of values of $\Delta z_i = z(C~IV)_i - z
(H~I)_i$. We have already seen that 
$b(C~IV)_i \simeq 1/2~b(H~I)_i$, 
indicative of the fact that \lya\ absorption
takes place over a wider velocity range than \CIV. If we compare 
the {\it measured} values of $z(C~IV)_i$ and  $z(
H~I)_i$ in the 20 \lya\ clouds with log~$N$(H~I) $\geq 14.5$
which show C~IV absorption (\S 4 above), we find a distribution
of values of $\Delta z_i$ with $\sigma_z = 4 \times 10^{-4}$
which corresponds to a velocity dispersion of 27~\kms. We
therefore repeated the simulations above assuming that the
C~IV$\lambda 1548$ lines to be stacked are drawn at random from a
Gaussian distribution of $\Delta z_i$ with this value of $\sigma$; the
results are reproduced in Figures 6b, c, and d. 

Clearly, once this redshift dispersion is introduced in the
sample, the signal in the stacked spectrum is blurred to the
point where it is questionable whether it would be detected even
if $N$(C~IV)/$N$(H~I) $ = 2.5 \times 10^{-3}$. Although the
average residual intensity in Figure 6b is less than one, the
absorption is so diffuse that it becomes difficult to distinguish
it from small fluctuations of the continuum level. 
We thus seem to have identified a fundamental limitation of the
stacking method for the detection of very weak absorption
features. Unless it can be shown that $\Delta z$ decreases with
log~$N$(H~I), it would appear that the uncertainty in the exact
redshift of the metal lines relative to \lya\ casts serious
doubt on the interpretation of non-detections in
composite spectra, even at a signal-to-noise ratio well in excess
of that of the observations reported here. In any case, this
probably explains why the first attempts at stacking QSO spectra
to search for weak C~IV lines (e.g. Lu 1991) underestimated the
typical $N$(C~IV)/$N$(H~I) ratio subsequently measured with HIRES
spectroscopy. It is interesting to note in this context that the
only case where the stacking method has produced an 
incontrovertible detection is when Barlow \& Tytler (1998)
co-added {\it HST} FOS spectra to detect C~IV associated with
\lya\ clouds at low redshift ($\langle z \rangle \simeq 0.5$). At
the coarse resolution of the FOS spectra---a factor of $> 25$
worse than that of HIRES data---the velocity difference between
C~IV and \lya\ becomes a secondary effect.

\subsection{Pixel-by-pixel Comparison}

Cowie \& Songaila (1998) have recently proposed a novel way to look for 
weak absorption features associated with \lya\ forest lines. The method
involves constructing cumulative distributions of optical depths for all 
the pixels in a spectrum where (in this case) C~IV~$\lambda 1548$
absorption may be found, that is all pixels at wavelengths 

\begin{equation}
	\lambda_j = \frac{1548.195}{1215.670} \times \lambda_i
	\label{}
\end{equation}

\noindent where $\lambda_i$ is the wavelength of 
the $i$th pixel in the \lya\ forest.
Since no C~IV absorption is expected for \lya\ pixels with 
optical depth $\tau {\rm (Ly\alpha)} < 0.1$, the distribution of 
corresponding values of $\tau {\rm (C~IV)}$ provides a reference `blank' 
sample. One can then examine the {\it difference} between the distribution
of $\tau {\rm (C~IV)}$ for a particular range 
of $\tau {\rm (Ly\alpha)}$ values
and the blank sample to determine whether a residual signal is 
present and indeed Cowie and Songaila (1998) reported excess 
C~IV absorption for all \lya\ optical depths, from 
$\tau {\rm (Ly\alpha)} = 20$ to 0.5\,.

We followed closely the procedure outlined by Cowie \& Songaila (1998)
in applying their method the spectrum of APM~08279+5525. The results 
are reproduced in Figure 7, where the left-hand panel shows the 
distributions of $\tau {\rm (C~IV)}$ for the same three ranges of 
$\tau {\rm (Ly\alpha)}$ considered by Cowie \& Songaila.
The differential distributions relative to the blank sample
are shown in the right-hand panel; here the deficit of pixels
with negative values of $\tau {\rm (C~IV)}$ and 
the excess of pixels with
positive $\tau {\rm (C~IV)}$ give an `S wave' pattern
which is indicative of C~IV absorption associated with \lya\ forest lines.
As can be seen from the Figure, we confirm the finding by Cowie \& Songaila
of a C~IV signal in all three intervals of 
$\tau {\rm (Ly\alpha)}$ considered.

It is of interest, of course, to ask how changes in the
$N$(C~IV)/$N$(H~I) ratio would be reflected in the optical depth 
distributions shown in Figure 7.
We are in a good position to explore this question having
quantified the absorption parameters of the entire \lya\ forest
in APM~08279+5525\footnote{Within the redshift limits designed to
exclude BAL material, as discussed at \S3 above}.
To this end, we simulated the C~IV absorption associated with 
375 \lya\ lines for which VPFIT returned values of 
$N$(H~I), $z$, and $b$, and applied Cowie and Songaila's technique to 
determine the distributions of $\tau {\rm (Ly\alpha)}$
and $\tau {\rm (C~IV)}$.
Since the aim is to test whether there is a sudden change in the 
metallicity of \lya\ clouds at low optical depths, we considered
two possibilities: 

\noindent (1) $N$(C~IV)/$N$(H~I)$ = 2.5 \times 10^{-3}$ 
for all values of $N$(H~I); and 

\noindent (2) a drop in $N$(C~IV)/$N$(H~I) by a factor of 10
for $N$(H~I)$ < 14.0$\,.

\noindent We carried out the simulations for two 
cases which we term `ideal' and 
`realistic'. 
In the ideal case, we did not include the effects of noise
(that is, we assumed infinitely high S/N) and we adopted
$b(C~IV)_i = b(H~I)_i$ and $z(C~IV)_i = z(H~I)_i$\,.
In the realistic case, we introduced random noise in the simulations
to reproduce the S/N = 80 typical of our HIRES data.
We also assumed $b(C~IV)_i = 1/2~b(H~I)_i$ and a dispersion of 
values $\Delta z_i = z(C~IV)_i - z(H~I)_i$ with 
$\sigma_z = 4 \times 10^{-4}$ as found for \lya\ lines 
with log~$N{\rm (H~I)} \geq 14.5$ (\S5.1)\,.
The results of these simulations are reproduced in Figure 8.
Since the assumed drop in the $N$(C~IV)/$N$(H~I) ratio 
at $N$(H~I)$ < 14.0$ (case 2)
would affect primarily pixels with $\tau$(\lya) $\simlt 1$, for
clarity we 
show the results at low optical depths in two subsets:
$ \tau$(\lya) $< 1$ and $1 \leq \tau$(\lya) $\leq  2$\,.

The top panel in Figure 8 shows that in the `ideal' case 
the optical depth method developed by Cowie and Songaila (1998) would 
indeed be sensitive to changes in the Carbon abundance of \lya\ clouds
(open squares).
Under more realistic conditions (bottom panel), however,
the break below $\tau$(\lya)$ \simeq 1$ is softened, primarily 
by the effect of noise which moves pixels with 
$1 \leq \tau$(\lya) $\leq  2$ into the lower optical depth
interval and {\it vice versa}. Furthermore, at the typical S/N of
the present data, the error associated with the lowest
$\tau$(C~IV) bin is so large that the two cases considered
(filled dot and open square) 
can no longer be distinguished with confidence.
Although the data (open star) apparently favour a
constant $N$(C~IV)/$N$(H~I), we feel that the existence of a
break below $\tau$(\lya)$\simeq 1$ 
can only be tested reliably with higher S/N observations.

\section{Summary and Conclusions}

We have presented a high resolution ($\sim$ 6 \kms) and high 
S/N ($\sim 80$) spectrum of the ultraluminous BAL QSO
APM~08279+5255 obtained with HIRES on the Keck I telescope. These
data, which have a sensitivity to C~IV $\lambda\lambda 1548,1550$
lines with rest frame equivalent widths as low as $\sim$3~m\AA\
($5\sigma$), have been analyzed with a view to reassessing the
metallicity of the \lya\ forest at $z \simeq 3.4$. Our principal
results are as follows.

In agreement with previous analyses, we find that approximately 
50\% of \lya\ clouds with log~$N$(H~I)$ \geq 14.5$ have
associated C~IV systems with log~$N$(C~IV)$ \simgt 12.0$; we
deduce a median $N$(\CIV)/$N$(\HI)$\simeq 1.4 \times 10^{-3}$.
However, we point out that the criterion previously adopted to
identify such \lya\ clouds---a residual flux in the line core
$r_f < 0.025$---is an inadequate approximation which can miss a
significant proportion of the lines. Even though the absorptions
are close to saturation, profile fitting methods which
simultaneously determine the column density and velocity
dispersion of the absorbers are preferable for a clean definition
of the sample (for data of sufficiently high S/N and resolution,
such as those presented here).

We have stacked the \CIV $\lambda 1548$ regions corresponding to 51 \lya\ lines
with $13.5 \leq$ log~$N$(H~I) $\leq 14.0$ but find no detectable
signal in the composite spectrum. In order to understand the
significance of this null result, we have performed simulations 
in which we stack 51 synthetic \CIV$\lambda 1548$ lines with
absorption parameters resembling as closely as possible those
of the lines we are trying to detect. Specifically, each \CIV\
line to be stacked was assigned values of column density and velocity
dispersion scaled directly from those of the
corresponding \lya\ line. These simulations show that we should
be able to detect a marginally significant ($4 \sigma$)
absorption feature in the co-added spectrum if 
$N$(C~IV)/$N$(H~I)$ = 1.4 \times 10^{-3}$, as in clouds with
log~$N$(H~I) $\geq 14.5$, {\it provided that there is no random
difference $\Delta z$ between the redshifts of C~IV and \lya.}

However, when such a difference is introduced in the simulations
the co-added signal is effectively washed out to the point where
it can no longer be recognized. This is the case even for a 
relatively small dispersion of values of $\Delta z$ ($\sigma = 
27$~\kms) which is entirely plausible given the {\it
observed} distribution of $\Delta z$ in the higher column density
clouds where \CIV is detected. We propose that this uncertainty
in the exact registration of weak signals is a serious problem
which ultimately limits the usefulness of stacking procedures.

We also analysed our data with the pixel-by-pixel optical depth method 
recently developed by Cowie \& Songaila (1998) and, in agreement with 
these authors, we do find a signal indicative of C~IV absorption 
in \lya\ clouds with optical depths as low as 
$\tau {\rm (Ly\alpha)} = 0.5 - 2$\,. 
However, our simulations show that a higher S/N than achieved 
up to now is required to detect with confidence even a marked break in
the $N$(C~IV)/$N$(H~I) ratio below log~$N$(H~I)$\simlt 14.0$\,.

We conclude that the question of whether the abundance of Carbon
in the \lya\ forest is uniform at all column densities has yet to be
fully answered. Probably this question is best addressed by
pushing the detection limit for 
\CIV\ absorption
lines below the limits reached so far. Thanks to the
extraordinary luminosity of APM~0827+5255 this approach is now
feasible with a concerted effort of HIRES observations.

\section{Acknowledgements}

The authors would like to express their gratitude to both Craig Foltz
and Michael Rauch for obtaining some of the observations presented in
this paper.  We are also grateful to Tom Barlow for his help using HAR
and to Bob Carswell, Jim Lewis and Jon Willis for their generous help
with various stages of data reduction.
SLE would like to thank the University of Victoria and the University
of Washington for hosting visits and especially Arif Babul and Chris Pritchet
at the University of Victoria for financial support.  SLE acknowledges
PPARC for her PhD studentship and WLWS acknowledges support 
from NSF Grant AST-9529073.

\newpage

\begin{figure}
\figurenum{1}
\epsscale{0.7}
\plotone{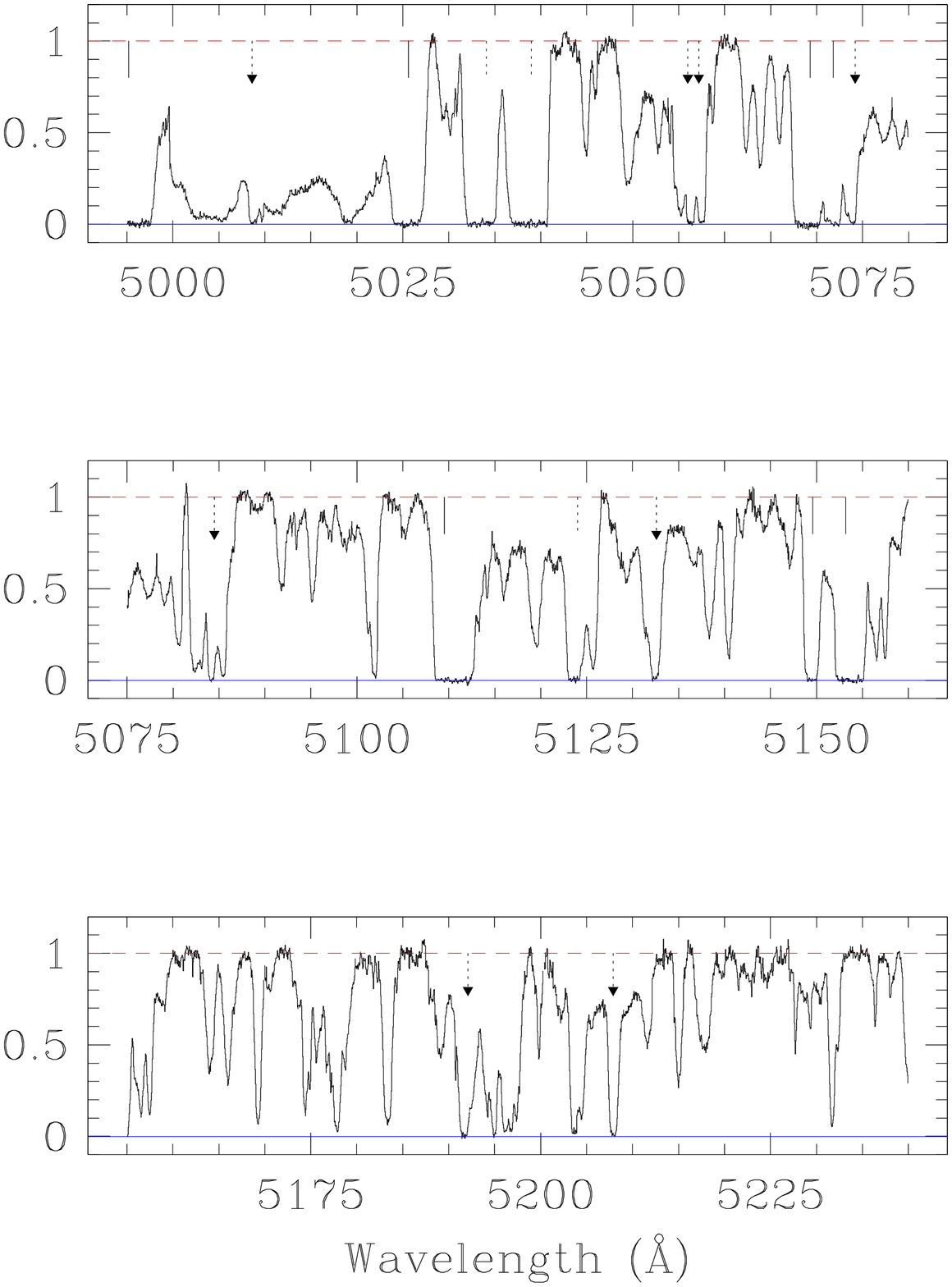}
\caption{The \lya\ forest in APM~08279+5255 within the redshift
range $z = 3.109 - 3.701$ of our analysis. The $y$-axis is
normalized counts. Vertical tick marks near the continuum level
indicate \lya\ lines with residual flux $r_f < 0.025$; solid and
broken tick marks are used to indicate lines with and without
associated C~IV~$\lambda\lambda 1548, 1550$ absorption,
respectively. Solid triangles flag \lya\ lines with column
density log~$N$(H~I)$< 14.5$ even though $r_f < 0.025$ (see
text). The horizontal black line between 5387 and 5456~\AA\
indicates the region of the forest omitted from the analysis
because the corresponding C~IV wavelengths fall within the
atmospheric B band.}
\end{figure}

\newpage

\begin{figure}
\figurenum{1}
\epsscale{0.9}
\plotone{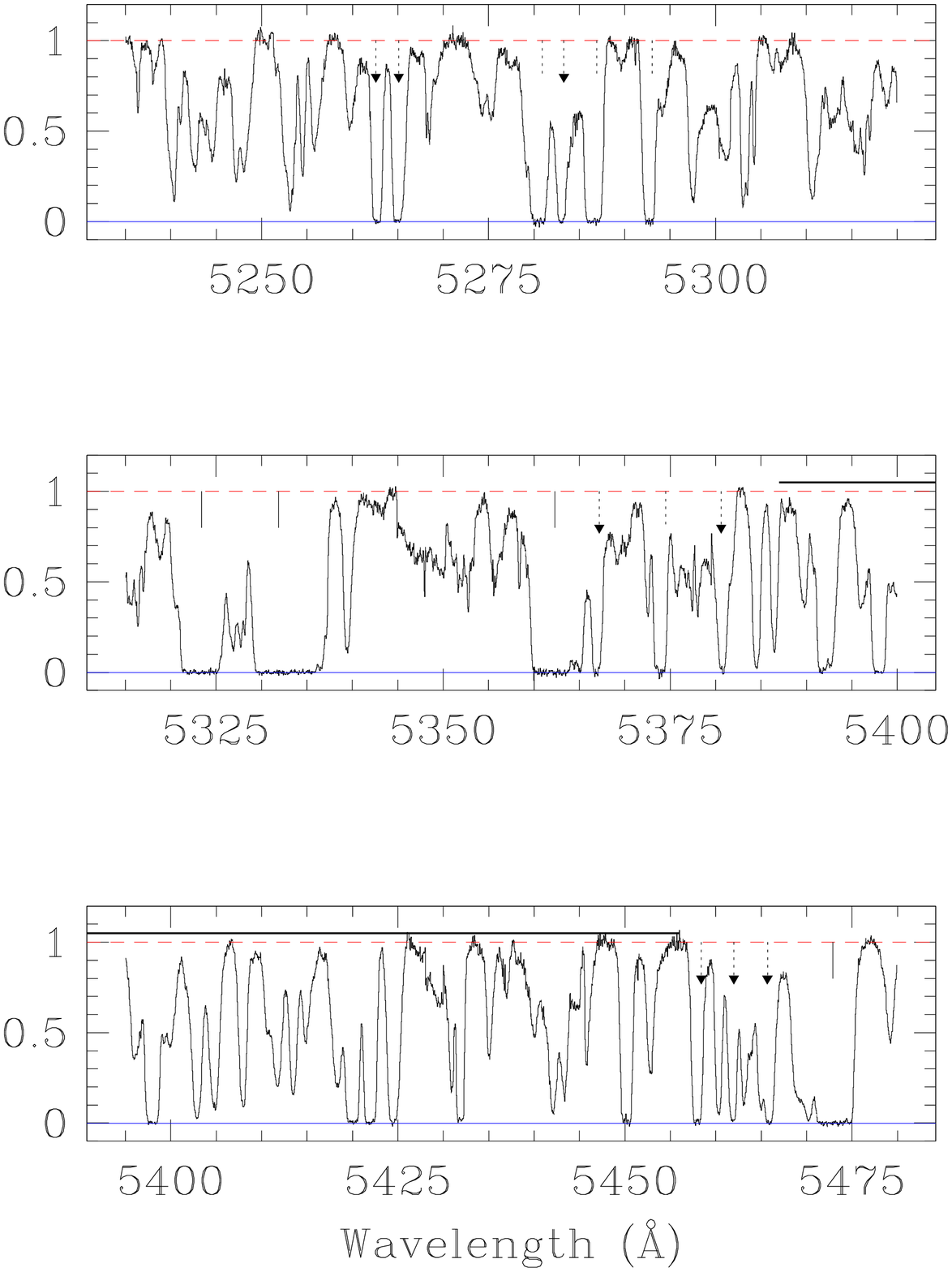}
\caption{Continued}
\end{figure}

\newpage

\begin{figure}
\figurenum{1}
\epsscale{0.9}
\plotone{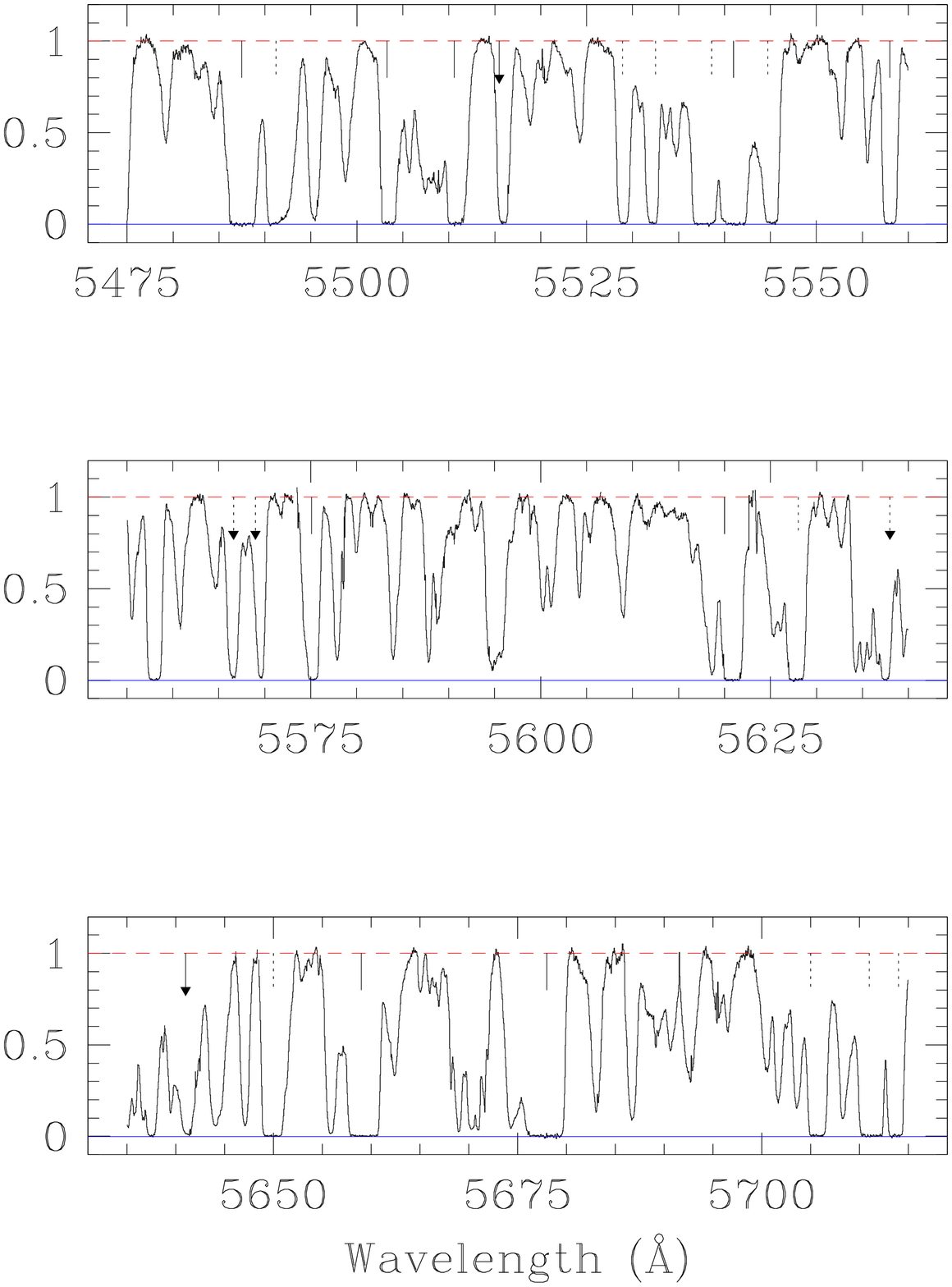}
\caption{Continued}
\end{figure}
\newpage

\begin{figure}
\figurenum{2}
\epsscale{0.8}
\plotone{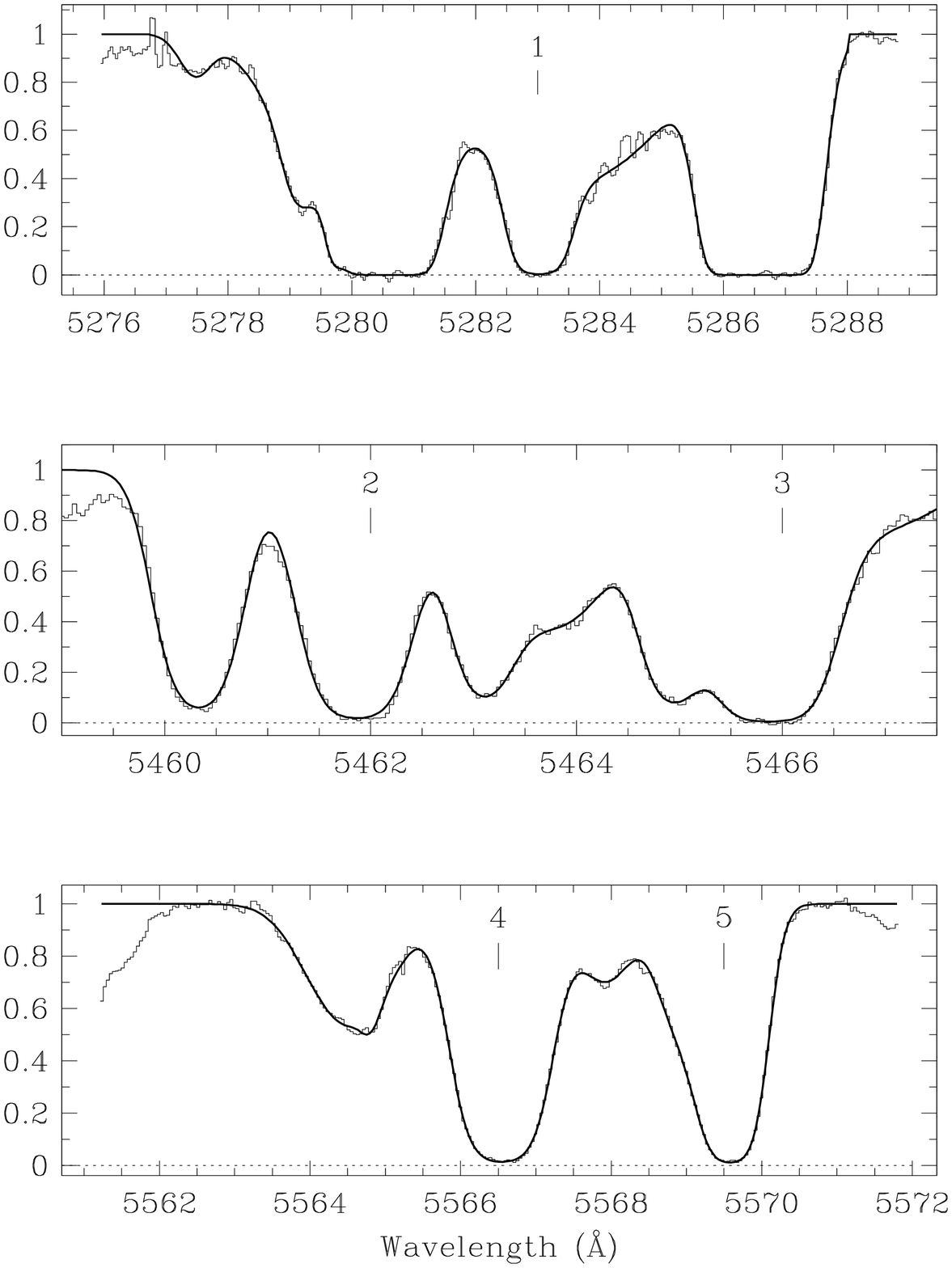}
\caption{Examples of \lya\ absorption lines with $r_f < 0.025$
and log~$N$(H~I)$< 14.5$. Theoretical line profiles computed with
VPFIT are shown as a continuous line in each panel. Parameters of
the fits are collected in Table 2.} 
\end{figure}

\newpage
\begin{figure}
\figurenum{3}
\epsscale{0.7}
\plotone{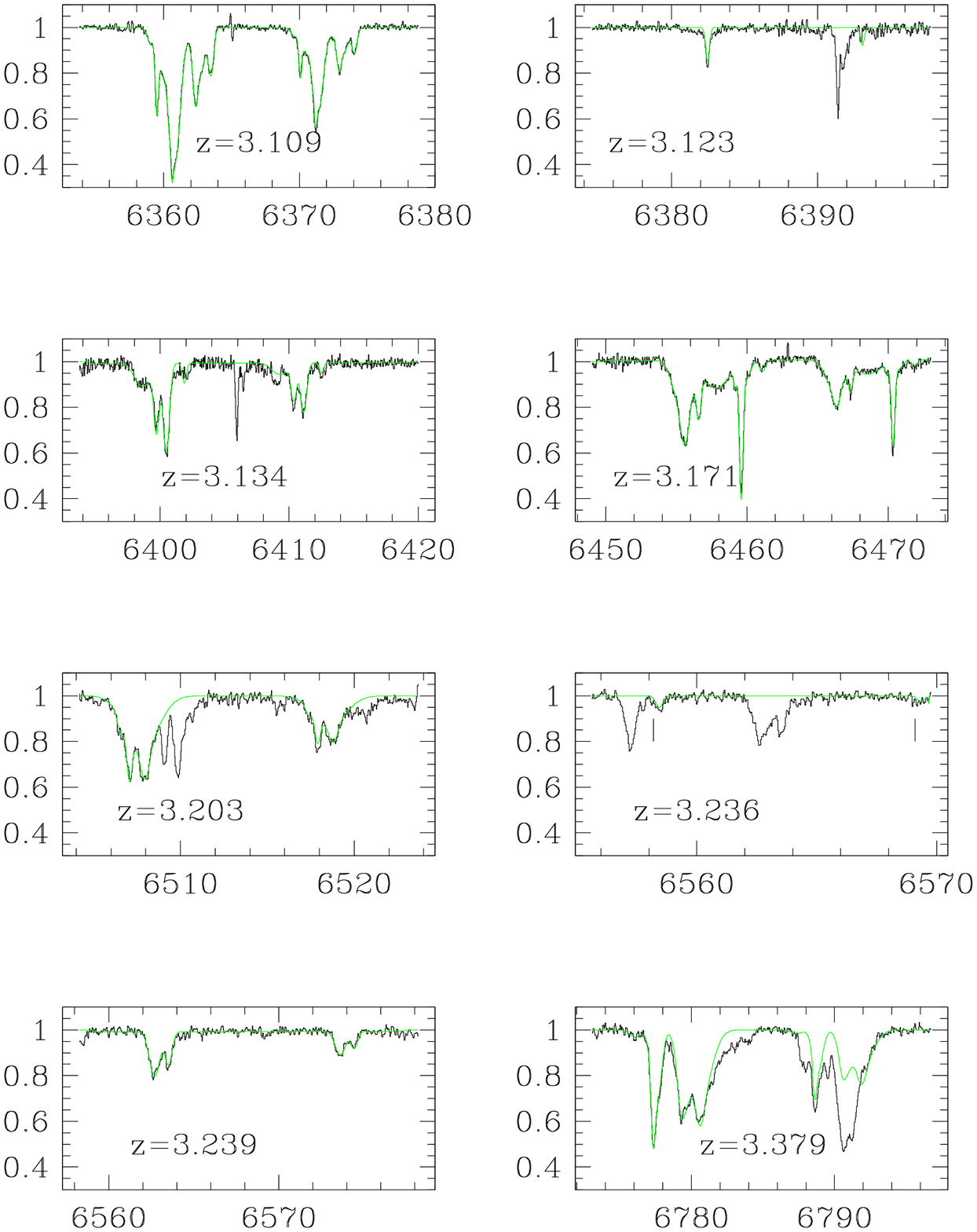}
\caption{Atlas of C~IV doublets associated with \lya\ lines in
our working region. The $x$-axis is wavelength in \AA\ and the $y
$-axis is normalized counts. Thin dotted lines show the
profile fits with the parameters listed in Table 2.
The weakest C~IV systems are indicated with tick marks to guide
the eye. }
\end{figure}

\newpage

\begin{figure}
\figurenum{3}
\epsscale{0.9}
\plotone{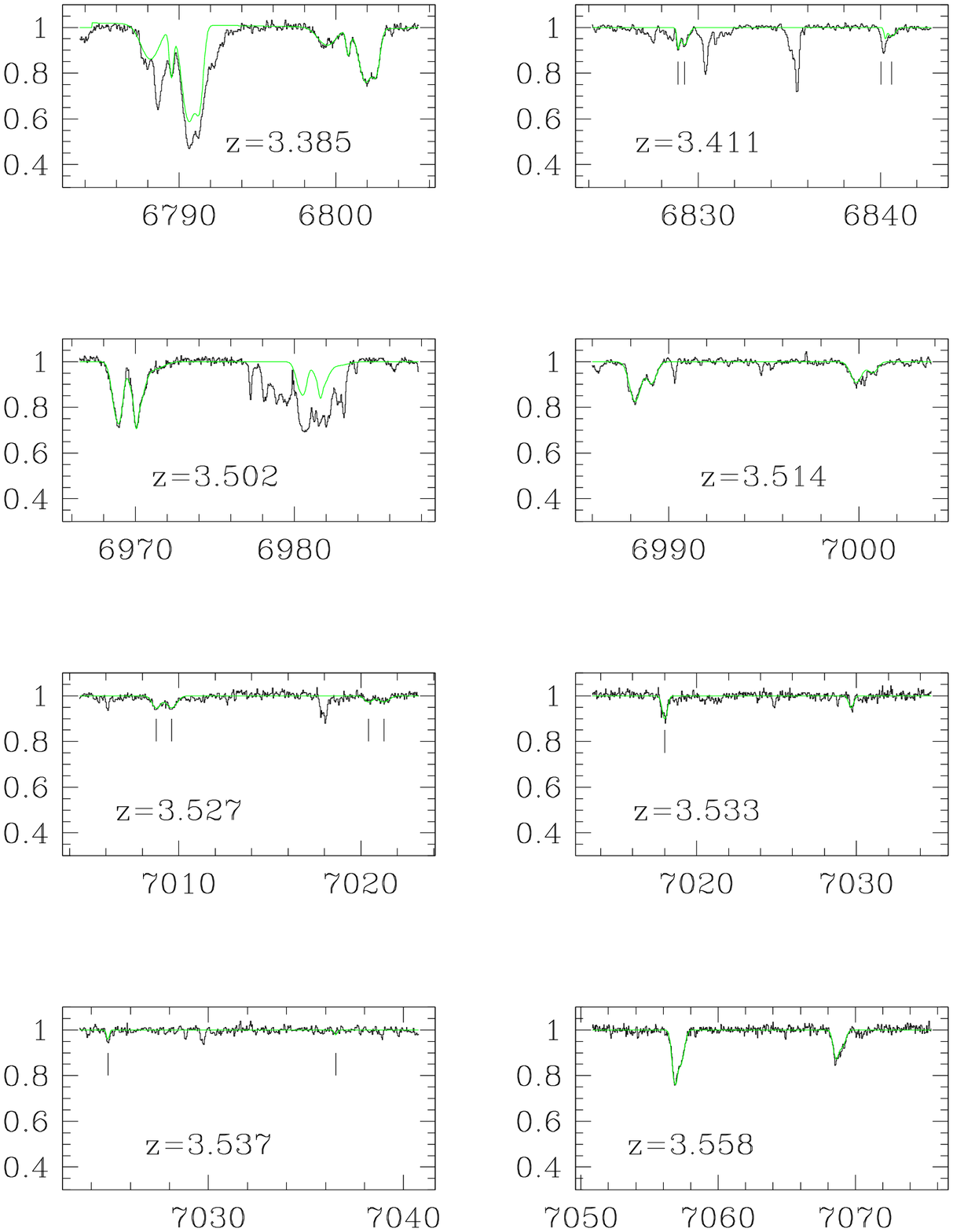}
\caption{Continued}
\end{figure}

\newpage

\begin{figure}
\figurenum{3}
\epsscale{0.9}
\plotone{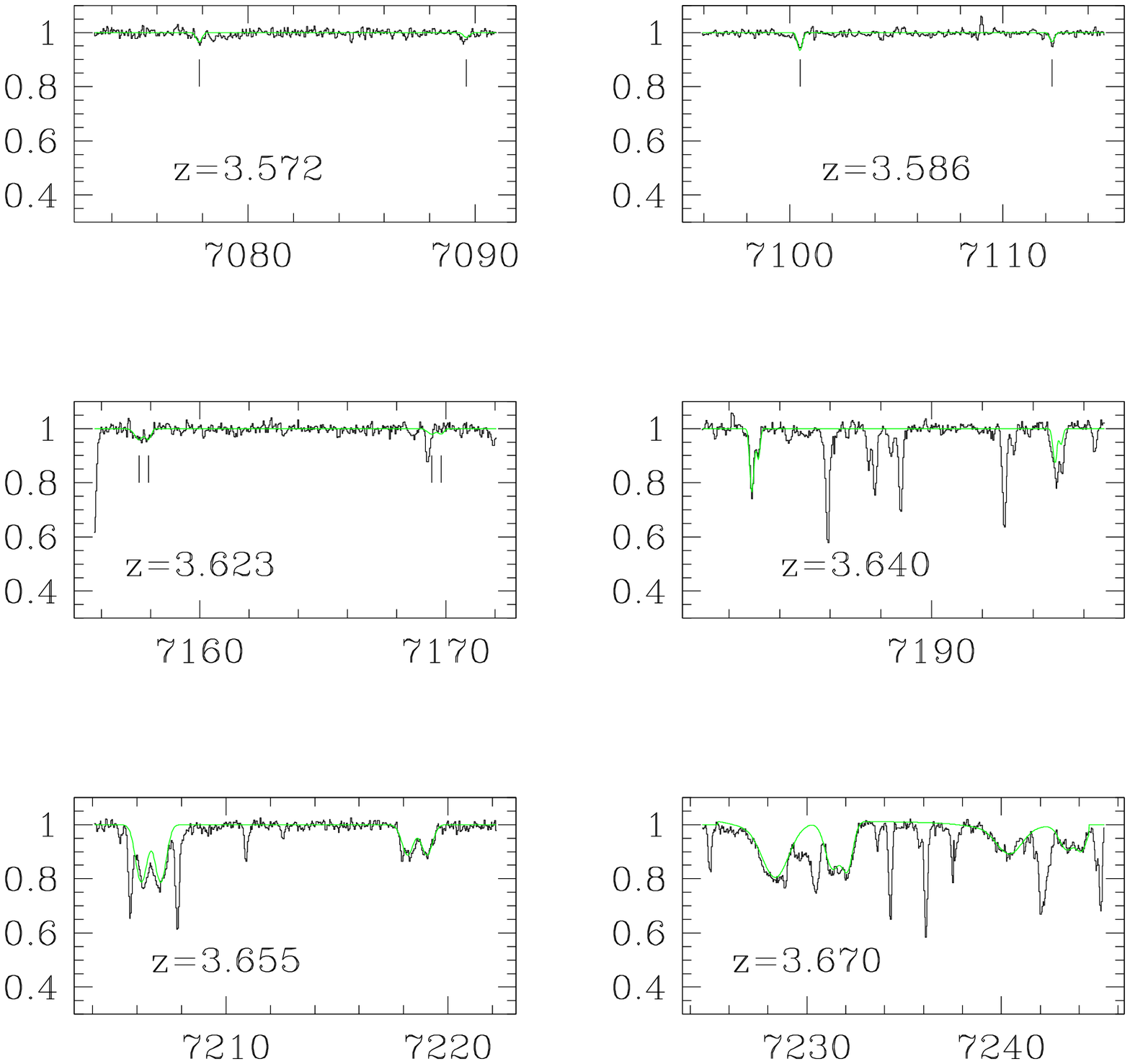}
\caption{Continued}
\end{figure}


\newpage

\begin{figure}
\figurenum{4}
\hspace*{1.5cm}
\psfig{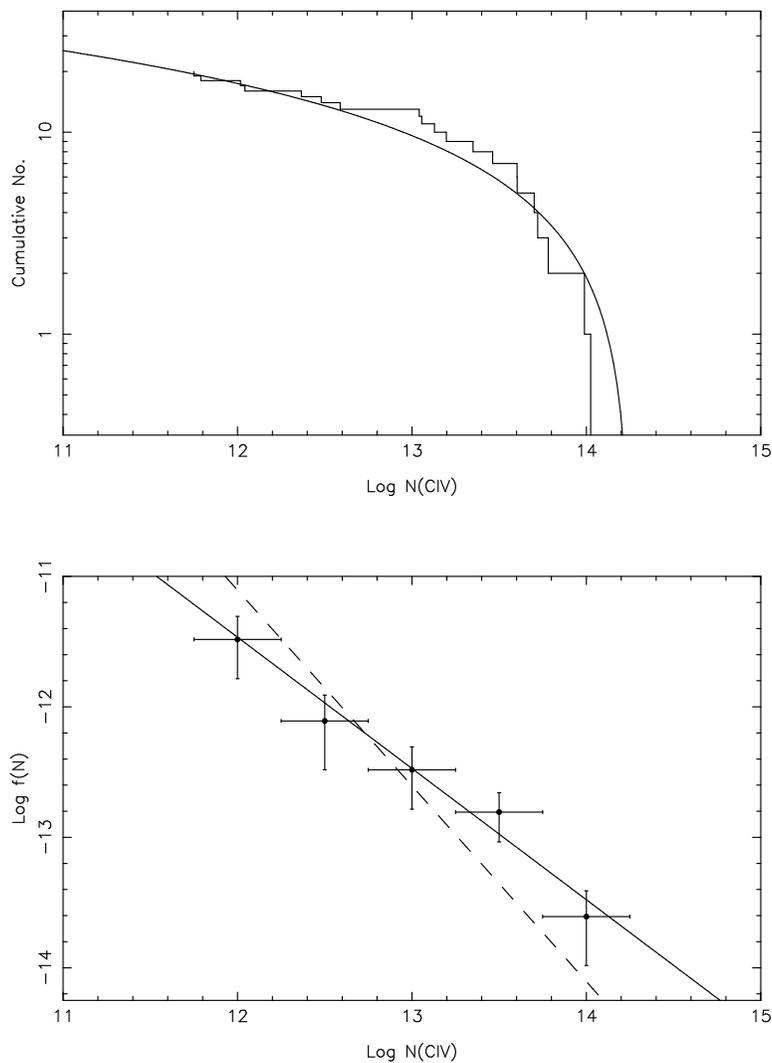}
\caption{Column density distribution of C~IV absorbers in
APM~08279+5255. The data have been grouped into bins of width 0.5
in log~$N$(C~IV) for display purposes only. The top panel shows
the cumulative distribution and the bottom panel is the
differential.  A maximum likelihood
analysis of the distribution, assumed to be a power-law of the
form $f(N) dN = BN^{\alpha} dN$, returned a best fit index
$\alpha = -1.0$ indicated by the solid line. The dashed line
in the bottom panel ($\alpha = -1.5$) is the slope reported
by Songaila (1997) for her larger sample.}
\end{figure}

\newpage

\begin{figure}
\figurenum{5}
\hspace*{-1.25cm}
\psfig{figure=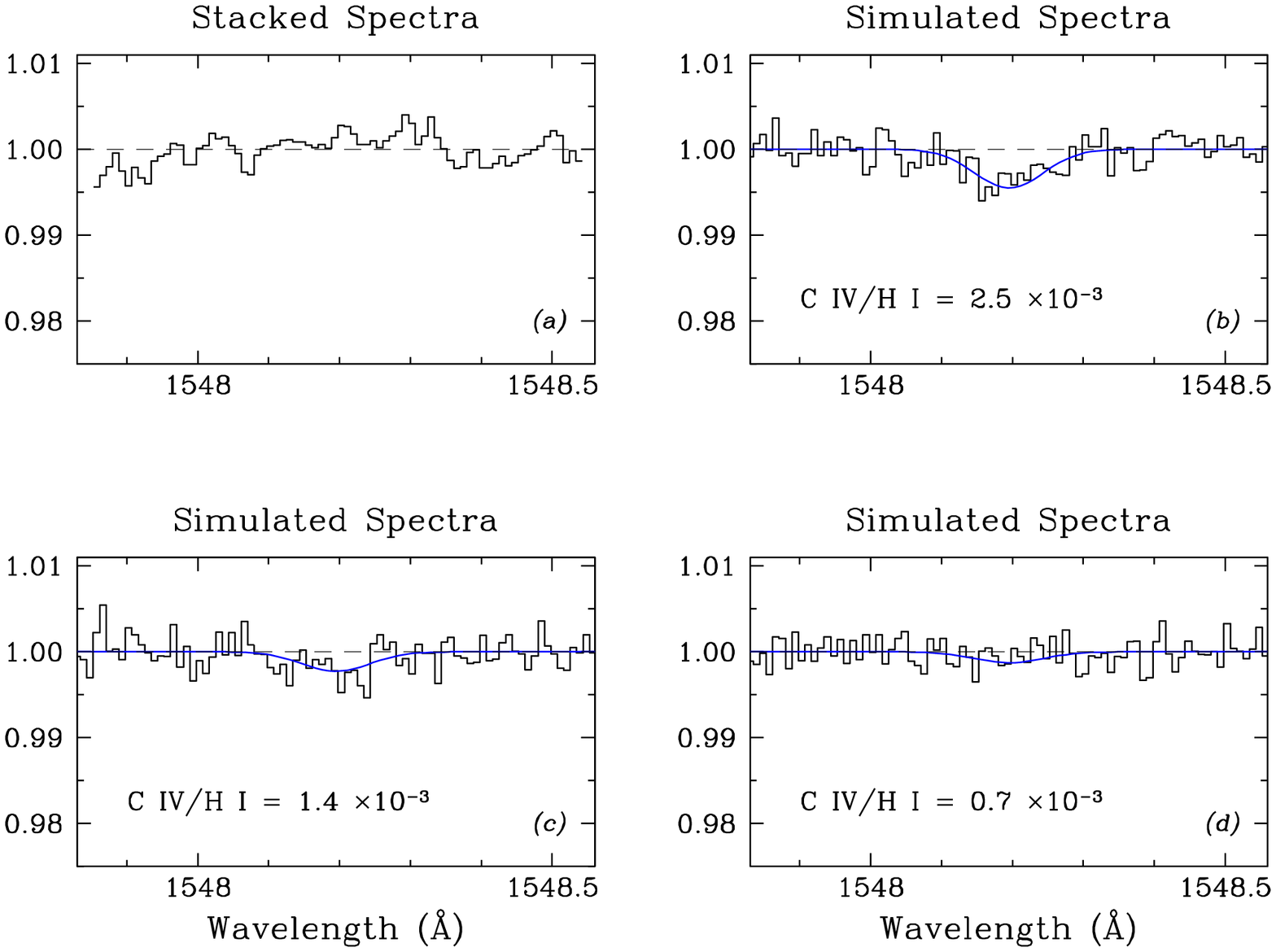,width=185mm,angle=270}
\caption{{\it (a)}: Co-added spectrum obtained by stacking 51 C~IV
regions corresponding to \lya\ lines with $13.5 \leq$
log~$N$(H~I) $\leq 14.0$\,. The resulting S/N is 580. {\it (b),
(c), {\rm and} (d)}: Simulated stacked spectra for different
values of $N$(C~IV)/$N$(H~I), as indicated. The continuous line
in each panel shows the noise-free absorption profile for a
single C~IV line with the mean $N$(C~IV). These simulations
assumed no redshift difference between C~IV and \lya\ lines.}
\end{figure}

\newpage

\begin{figure}
\figurenum{6}
\hspace*{-1.25cm}
\psfig{figure=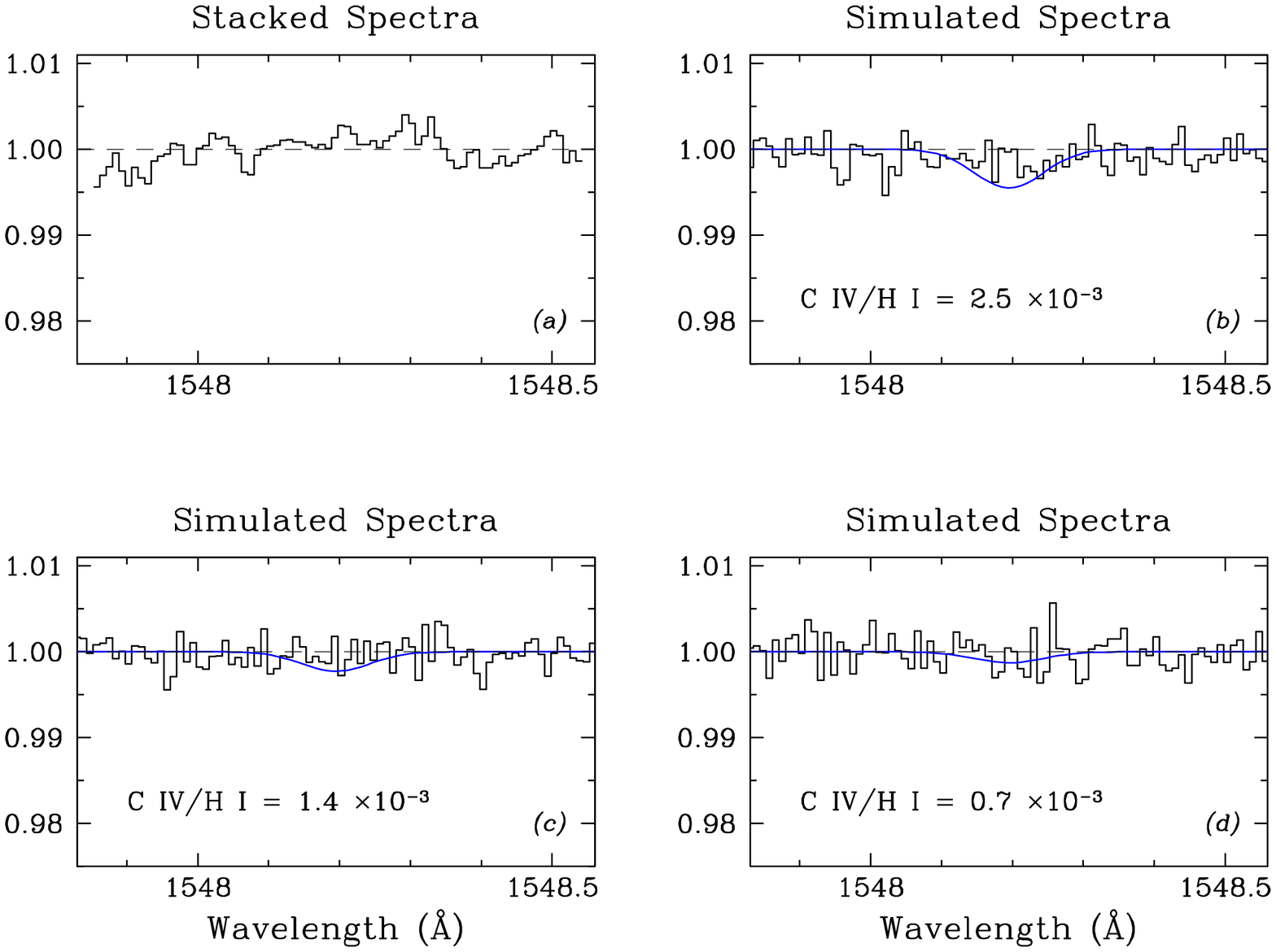,width=185mm,angle=270}
\caption{As Figure 5, except that the simulated spectra include
allowance for a random redshift difference $\Delta z$ between the
centroids of C~IV and \lya\ absorption. Values of $\Delta z$ were
drawn from a Gaussian distribution with $\sigma = 27$~km~s$^{-1}$.}
\end{figure}

\newpage

\begin{figure}
\figurenum{7}
\hspace*{0.5cm}
\psfig{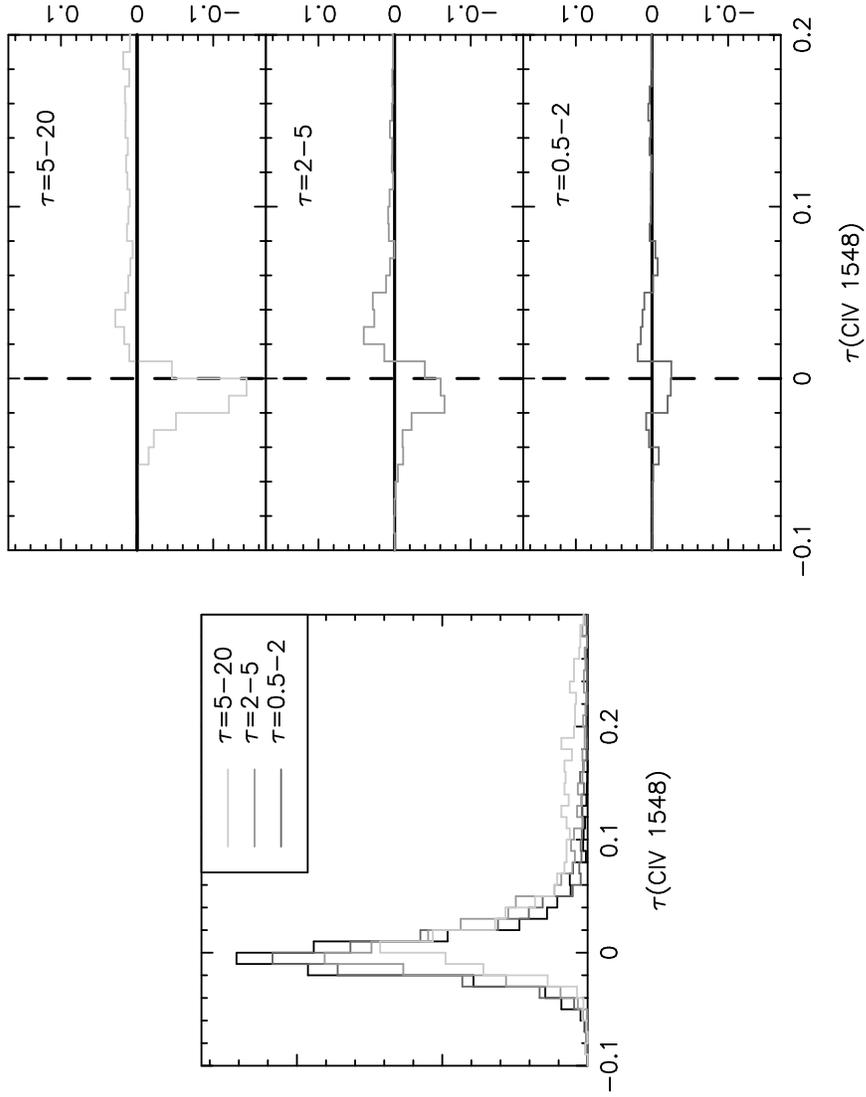}
\caption{{\it Left:} Distribution of C~IV optical depths for different
ranges of \lya\ optical depth, as indicated.~ 
{\it Right:} Differential
distributions of C~IV optical depths, relative to that for 
$\tau {\rm (Ly\alpha)} < 0.1$ where no C~IV absorption is expected.}
\end{figure}

\newpage

\begin{figure}
\figurenum{8}
\vspace*{-1cm}
\psfig{figure=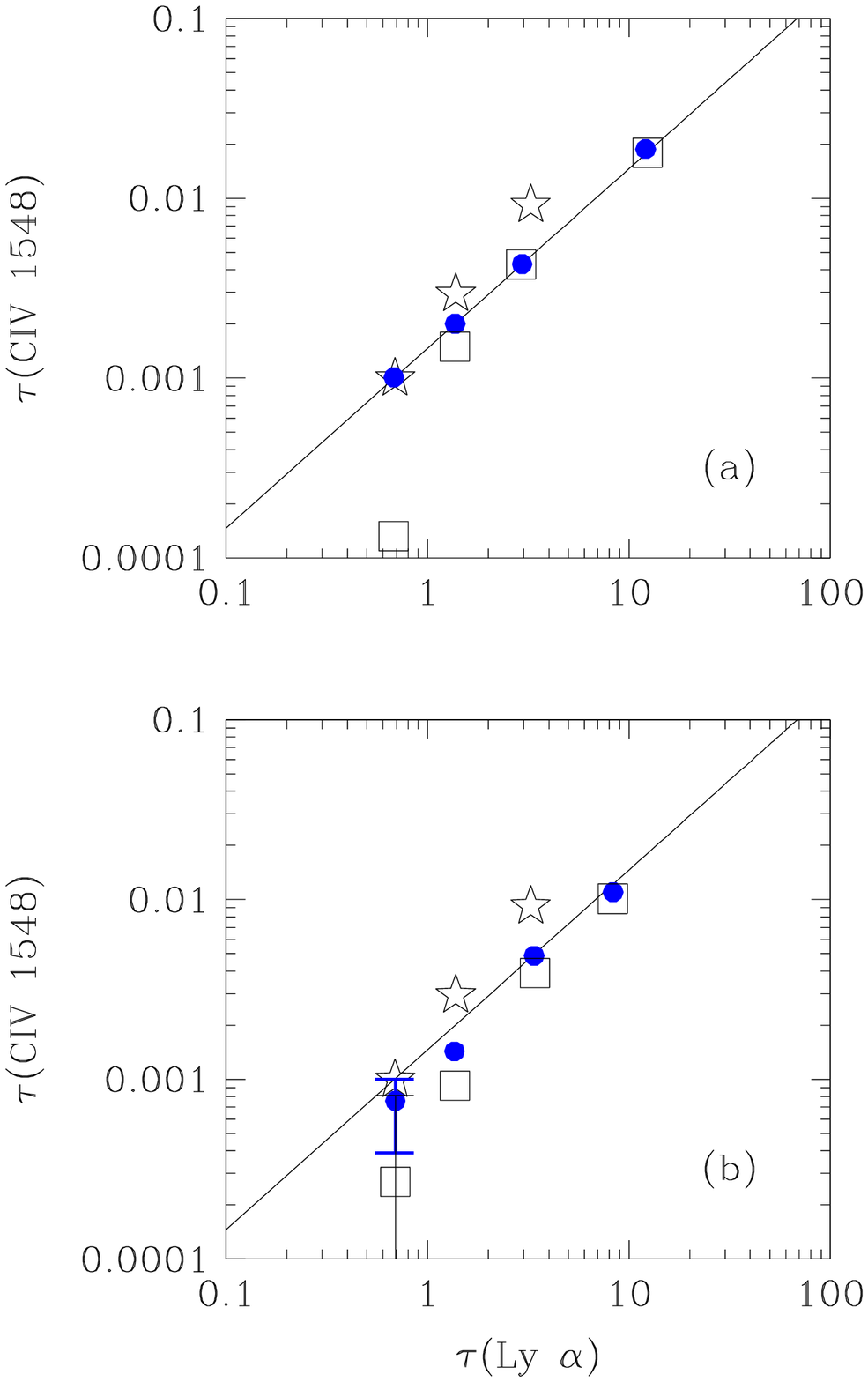,width=4.5in}
\vspace{-2.75cm}
\caption{Median values of optical depth of C~IV~$\lambda 1548$
for different ranges of \lya\ optical depth. 
The open star symbols are the values measured in the HIRES spectrum
of APM~08279+5525; no value is plotted for the bin
$\tau$(\lya)$= 5 - 20$ because the residual intensity in these 
pixels is too low to measure the optical depth without reference
to higher order Lyman lines.
The filled circles and open squares show the results of 
the simulations described in the text.
The filled circles are for
the case where $N$(C~IV)/$N$(H~I)$= 2.5 \times 10^{-3}$
(straight line) was assumed throughout, whereas the open squares
are for simulations where a sudden drop by a factor of 10 in 
$N$(C~IV)/$N$(H~I) was introduced at log~$N$(H~I)$ < 14.0$\,.
The top and bottom panels correspond respectively to the `ideal' and
`realistic' simulations discussed in the text.
In the lower panel the error bars (only plotted if they are
larger than the symbols) show the dispersion 
in the results when the same simulation was repeated
twelve times.
}
\end{figure}

\newpage

\begin{deluxetable}{cccc}
\footnotesize \tablecaption{
Journal of HIRES Observations of APM 08279+525}
\tablewidth{400pt}
\tablehead{
\colhead{Date} & 
\colhead{Integration time (s)} &  
\colhead{Wavelength range (\AA)} &
\colhead{Typical S/N}
}
\startdata
April 1998 & 1800 & 4400 -- 5945 & 15 \nl
April 1998 & 1800 & 4400 -- 5945 & 15 \nl
April 1998 & 1800 & 4400 -- 5945 & 15 \nl
April 1998 & 1800 & 4400 -- 5945 & 15 \nl
May 1998   & 2700 & 4410 -- 5950 & 25 \nl
May 1998   & 2700 & 4400 -- 5945 & 25 \nl
May 1998   & 900  & 5440 -- 7900 & 30 \nl
May 1998   & 3000 & 5440 -- 7900 & 60 \nl
May 1998   & 3000 & 5440 -- 7900 & 60 \nl
May 1998   & 3000 & 5475 -- 7830 & 55 \nl
May 1998   & 3000 & 5475 -- 7830 & 55 \nl
May 1998   & 3000 & 6765 -- 9150 & 50 \nl
May 1998   & 3000 & 6850 -- 9250 & 50 \nl
& & & \nl
Summed Total & 31\,500 & 4400 -- 9250 & 30 --150 \nl
\enddata
\end{deluxetable}

\begin{deluxetable}{ccccl}
\footnotesize \tablecaption{
Details of Absorption Components Fitted to Each C~\textsc{IV} System.} 
\tablewidth{350pt}
\tablehead{
\colhead{System No.} &
\colhead{Redshift} &
\colhead{log~$N$(C~\textsc{IV})} &
\colhead{$b$ (km s$^{-1}$)} &
\colhead{Comments}
}
\startdata
C1 & 3.10742 & 12.50 & 17.2 & \nl
   & 3.10769 & 12.89 & 6.7 & \nl
   & 3.10791 & 12.62 & 10.5 &\nl
   & 3.10841 & 12.50 & 3.7 &\nl
   & 3.10850 & 13.74 & 25.7 &\nl
   & 3.10952 & 12.67 &  9.2 &\nl
   & 3.10966 & 13.21 & 28.2 & \nl
   & 3.11027 & 12.78 &  11.8 &\nl
C2 & 3.12252 & 12.50 & 7.8 & log~$N$(\HI)$<$ 14.5 \nl
C3 & 3.13296 & 12.98 & 39.0 &\nl
   & 3.13363 & 12.30 & 22.5 &\nl
   & 3.13367 & 12.97 & 12.4 &\nl
   & 3.13416 & 13.23 & 14.9 &\nl
   & 3.13505 & 12.33 & 9.6 & \nl
C4 & 3.16966 & 13.35 & 38.5 &\nl
   & 3.16976 & 12.90 & 16.3 &\nl
   & 3.17039 & 12.69 & 9.8 & \nl
   & 3.17117 & 13.24 & 58.6 & \nl
C5 & 3.17205 & 12.13 & 5.7 &\nl
   & 3.17233 & 13.18 & 6.7 & \nl
   & 3.17237 & 12.34 & 15.8 &\nl
   & 3.17264 & 12.34 & 15.8 &\nl
   & 3.17330 & 12.19 & 13.3 &\nl
C6 & 3.20266 & 12.23 & 9.9 &\nl
   & 3.20300 & 12.76 & 8.8 & Some of 1548 \AA\ lines blended\nl
   & 3.20345 & 13.55 & 58.9 &  \nl
   & 3.20359 & 12.84 & 15.8 &\nl
C7 & 3.23618 & 12.04 &8.4 &  \nl
C8 & 3.23895 & 12.98 & 19.4 & \nl 
   & 3.23950 & 12.58 & 10.5 & \nl 
C9 & 3.37677 & 12.24 & 24.5 & Some blending in both doublet lines \nl
   & 3.37757 & 12.81 & 6.5 & \nl
   & 3.37770 & 13.28 & 19.7 & \nl
   & 3.37882 & 12.37 & 10.8 & \nl
   & 3.37891 & 13.24 & 22.5 & \nl
   & 3.37969 & 13.56 & 45.4 & \nl
   & 3.37969 & 12.89 & 17.0 & \nl
C10& 3.38456 & 13.13 & 36.0 & 1548 \AA\ line is blended with the 1550
\AA\ \nl
   & 3.38543 &12.65  &7.4 & line from the previous system.\nl
   & 3.38616 & 13.51 & 25.9 &\nl
   & 3.38660 & 13.02 & 13.4 & \nl
C11& 3.41088 & 11.90 & 1.5&1550 \AA\ line is blended.  \nl
   & 3.41111 & 12.35 &11.9&\nl
C12& 3.50125 &  13.01 & 16.8 &Some blending in both doublet lines \nl
   & 3.50141 & 12.39 & 9.9 & \nl
   & 3.50204 & 12.24 & 5.3 & \nl
   & 3.50209 & 13.08 & 22.6 & \nl
   & 3.50281 & 12.39 &  32.8 & \nl
C13& 3.51380 & 12.88 & 17.1 & \nl
   & 3.51436 & 12.54 & 14.8 &\nl
C14& 3.52706 &  12.24 & 13.1 &  \nl
   & 3.52760 & 12.33 & 16.9 & \nl
C15& 3.53303 & 12.37 & 8.9&  \nl
C16& 3.53744 & 11.63  & 2.4 & Very weak, 1550 \AA\ line barely visible\nl
C17& 3.55811 & 12.85 & 11.5 & \nl
   & 3.55842 & 12.63 & 12.7 &\nl
C18& 3.57168 & 11.79 & 6.6 & \nl
C19& 3.58630 & 12.02 &  5.5 & \nl 
C20& 3.62314 & 11.97 & 9.1 & \nl 
   & 3.62339 & 11.76 & 5.4 & \nl
C21& 3.63954 & 12.47 & 3.0 &1550 \AA\ line is blended \nl
   & 3.63968 & 12.09 & 1.9 & \nl 
C22& 3.65458 & 12.90 & 14.3 &1548 \AA\ line is blended  \nl
   & 3.65514 &12.89 &13.9 & \nl
C23& 3.66863 & 12.88 & 89.3 & Satisfactory fit, although slight blending\nl
   & 3.66892 & 13.22 & 36.9 &\nl
   & 3.67082 & 12.97 & 21.4 & \nl
   & 3.67131 & 12.81 & 15.3 &\nl
\enddata
\end{deluxetable}

\begin{deluxetable}{lccc}
\footnotesize \tablecaption{
Parameter Fits for Absorption Lines Shown in Figure 2}
\tablewidth{450pt}
\tablehead{
\colhead{Line No.} &
\colhead{Redshift} &
\colhead{log~$N$(\HI)} &
\colhead{$b$ (km s$^{-1}$)} 
}
\startdata
1 & 3.34577 & 14.18 & 23.6 \nl
2 & 3.49289 & 14.13 & 25.5 \nl
3 & 3.49618 & 14.28 & 27.8 \nl
4 & 3.57899 & 14.21 & 28.3 \nl
5 & 3.58146 & 14.08 & 22.7 \nl
\enddata
\end{deluxetable}

\end{document}